\documentclass[12pt,a4paper]{scrartcl}

\pdfoutput=1
\usepackage{FNAL}
\usepackage{definitions}

\title{Tests of Scintillator+WLS strips for Muon System at Future Colliders}

\date{\today}

\addauthor{Dmitri Denisov}{\institute{1}}
\addauthor{Valery Evdokimov}{\institute{3}}
\addauthor{Strahinja Lukić \thanks{slukic@vinca.rs}}{\institute{2}}

\addinstitute{2}{Vinča Institute, University of Belgrade, Serbia}
\addinstitute{3}{Institute for High Energy Physics, Protvino, Russia}
\addinstitute{1}{Fermilab, Batavia IL, USA}

\abstract{Prototype scintilator+WLS strips with SiPM readout for muon system at future colliders were tested for light yield, time resolution and position resolution. Depending on the configuration, light yield of up to 36 photoelectrons per muon per SiPM has been achieved, as well as time resolution of \besttime and position resolution of \bestpos.}

\graphicspath{ {./logos/}{./figures/} }

\begin{document}

\titlepage

\section{Introduction}
\label{sec:intro}

Several concepts of future colliders, including \epem colliders, are currently under study for the next generation of particle experiments \cite{LCC, FCCee, Cepc}. Due to the well-defined initial state of the interactions, low backgrounds and radiation levels, \epem colliders are an attractive option for precision measurements to test various theoretical extensions of the Standard Model in the areas where the predictions of the competing models differ by a few percent, such as, e.g.\ in the Higgs sector.

The detector concepts for the future \epem colliders have been developed to a high level of detail over the past decade. Since the publication of the LoI of the two major concepts, the SiD \cite{SiDLoI} and the ILD \cite{ILDLoI}, numerous technical details have been specified to an advanced level. R\&D prototypes of inividual subsystems reach levels of complexity involving hundreds of thousands of readout channels (See e.g.\ Ref.\ \cite{DHCAL-proto}).

However, for the muon systems relatively few specific details are defined. A promising option for the muon system consists of scintillator strips with WLS fibers and SiPM readout built into the magnet yoke of a detector in several layers \cite{ILCTDR4}. In such a system, the coordinates of the muon track are reconstructed using the observables such as the coordinate of the strip hit by a passing muon and the signal time difference between the ends of the strip to measure position along the strip.

The present study is the first in a series devoted to the study of the time resolution and the position resolution achievable from the time difference between the ends of scintillator strips with WLS fibers and SiPM readout. The measurements described in this paper have been performed using cosmic muons at the location of the \dzero assembly building at the Fermi National Accelerator Laboratory, Batavia, USA, at the elevation of 220~m above sea level. The local cosmic muon fluence has been measured previously by the MicroBooNE collaboration to be $\sim 100 \unit{m}^{-2} \unit{s}^{-1}$, with a peak energy between 1 and 2~GeV \cite{Woodruf14}. 

\hamamatsu SiPM were used for the test \cite{hamamatsu}.

Section \ref{sec:setup} describes the measurement setups that were used, Sec.\ \ref{sec:calibration} describes time- and amplitude calibration procedures, Sec.\ \ref{sec:configurations} describes the tested scintillator strip -- WLS fiber configurations, Sec.\ \ref{sec:analysis} gives details on the data analysis procedures, Sec.\ \ref{sec:results} tabulates the measurement results. Conclusions are given in Sec.\ \ref{sec:conclusions}.

\section{Measurement Setup}
\label{sec:setup}

Two setups that were used for the measurements are described here. In both setups cosmic muons are detected by coincidence between three vertically arranged scintillation counters. 

A CAMAC system with a LeCroy 2249A 12-input charge-sensitive ADC and a LeCroy 2228A 8-input TDC was used to digitize the amplitude and the arrival time of the signals. The data collection was performed and monitored from a PC with USB connection to the CAMAC Crate controller of type CC-USB by Wiener Plein\&Baus, using custom-made software \cite{WienerDAQ}.

\subsection{Setup 1}

\begin{figure}
\centering
\includegraphics{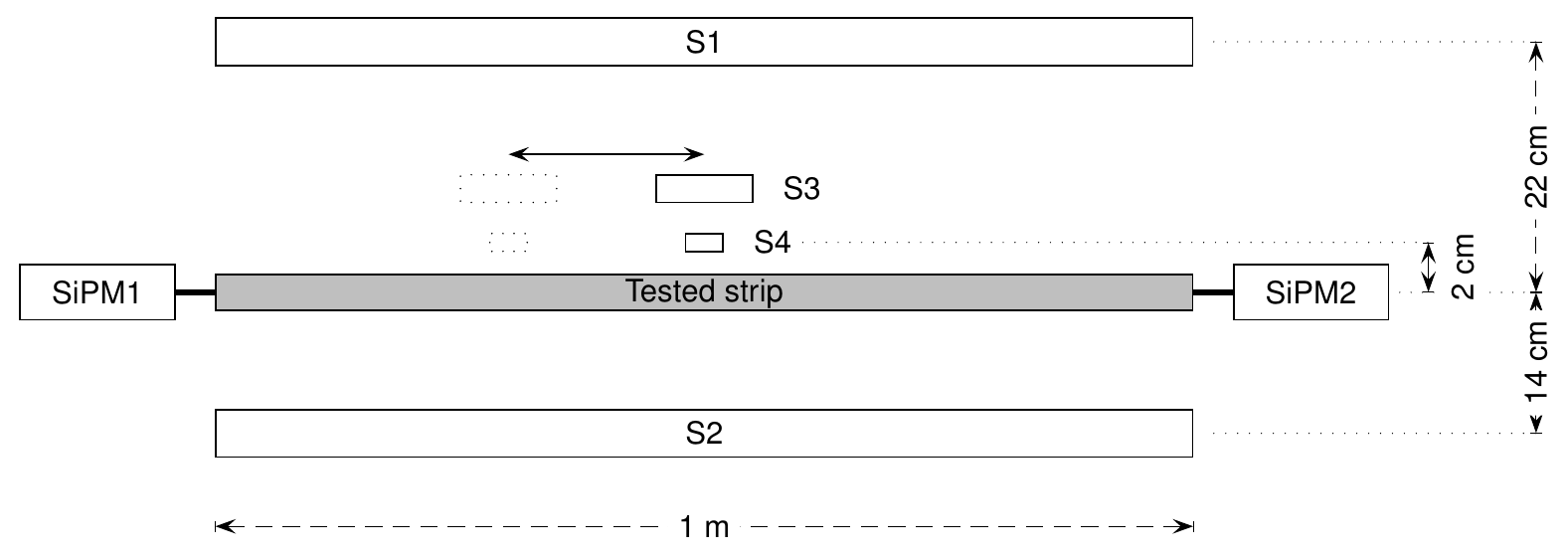}
\caption{\label{fig:setuppads}Schematic of the test setup with scintillator pads restricting cosmic muon position by coincidence. \sone and \stwo are scintillation counters with vacuum PMT positioned above and below the tested strip. \sthr and \sfour are small-area scintillators with vacuum PMT used to select events where the muon hits a given location along the tested strip. The location of \sthr and \sfour w.r.t the tested strip was changed from run to run, keeping the relative position of \sthr and \sfour always the same. \sipmo and \sipmt represent the SiPM connected to the respective ends of the WLS fiber of the tested strip.}
\end{figure}

Setup 1 is shown in Fig.\ \ref{fig:setuppads}. \sone and \stwo are plastic scintillation counters with vacuum PMT located above and below the tested strip, each of them 1~m long, 10~cm wide and 1~cm thick, used to detect the passage of a muon by coincidence requirement. \sthr and \sfour are scintillation counters with vacuum PMT used to restrict the location of the muon along the tested strip by coincidence requirement. \sthr is a $10\times15\unit{cm}$ plastic scintillator pad oriented with its 10~cm side along the tested strip. \sfour is a 40~cm long and 2.7~cm wide \bicron strip oriented across the tested strip. The distance from the center of the tested strip to the PMT of \sfour was 5~cm. 

The location of \sthr and \sfour along the tested strip was changed from run to run in order to study different points along the tested strip, keeping the relative position of \sthr and \sfour always the same. \sipmo and \sipmt denote the SiPM connected to the respective ends of the WLS fiber of the tested strip. The vertical distance between \sone and the tested prototype is 22~cm, and the vertical distance between \stwo and the tested prototype is 14~cm. The distance of \sfour to the tested prototype was 2~cm. The length of the tested strips was 1~m.

Coincidence between \sone, \stwo and \sthr was used as the trigger, signalling the passage of a muon. The signal from \sthr was delayed by 20~ns with respect to the signals from \sone and \stwo, so that the trigger signal is always formed at the rising edge of the \sthr signal. In the offline analysis, the presence of the signal in \sfour was required for event selection.

When measuring properties close to either end of the tested strip, the counters \sone and \stwo were moved along the axis in order to cover locations up to at least 20~cm beyond the end of the tested strip. This was done in order to prevent loss of the muon flux, which would have caused a loss of statistic and a muon position bias at these points.

The signals from the counters \sone and \stwo were used only for triggering. The signals from \sthr, \sfour, \sipmo and \sipmt were recorded. Each of the signals to be recorded was first split into the time- and the amplitude channels using linear fan-in fan-out modules. The time signals were processed using constant threshold discriminators. The time signals were delayed by $\sim 50\unit{ns}$ and digitized by the TDC CAMAC module using the trigger signal as the start. The amplitude signals were delayed by $\sim 30 - 40 \unit{ns}$ and digitized by the ADC CAMAC module, using gate generated by the trigger signal. 

\subsection{Setup 2}

\begin{figure}
\centering
\includegraphics{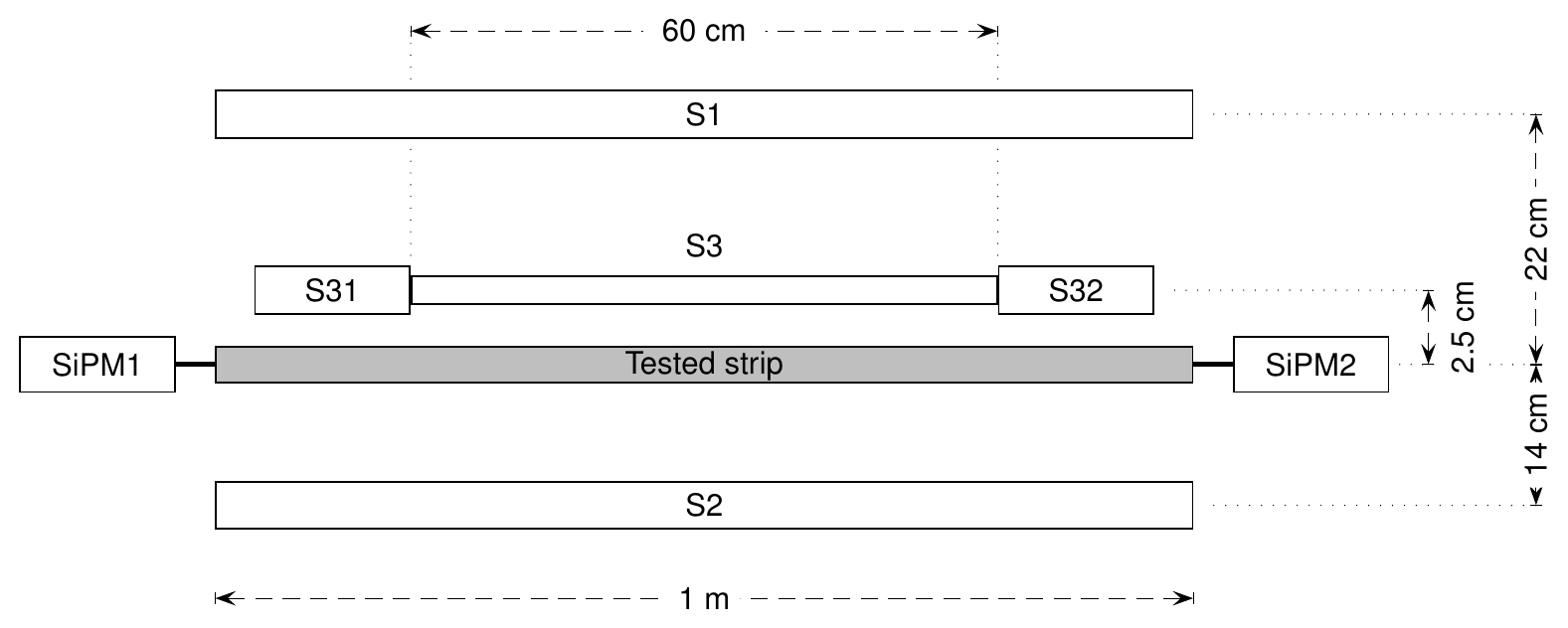}
\caption{\label{fig:setup}Schematic of the test setup with a scintillator strip with fast PMT for the muon position determination. \sone and \stwo are scintillation counters with vacuum PMT positioned above and under the tested strip. \sthro and \sthrt are vacuum PMTs attached to both ends of a scintillator strip located at a 2.5~cm distance vertically from the tested strip. \sipmo and \sipmt represent the SiPMs connected to the respective ends of the WLS fiber of the tested strip.}
\end{figure}

In setup 2 (Fig. \ref{fig:setup}) a \bicron scintillator strip, \sthr, was used to determine the location of the muon impact using the time difference between both ends readout by vacuum PMTs, \sthro and \sthrt. The length of \sthr is 60~cm, the width 2.7~cm and the thickness 1.2~cm. The counters \sone and \stwo are the same as in setup 1, located in the same way as in setup 1. 

Coincidence between \sone, \stwo and \sthro was used as the trigger, signalling the passage of a muon. The time signal of \sthro was delayed by 20~ns with respect to the signals of \sone and \stwo, so that the trigger signal is formed at the rising edge of the \sthro counter signal. 

The signals from \sone and \stwo were used only for triggering. The time and amplitude of the signals from \sthro, \sthrt, \sipmo and \sipmt were digitized in the same way as in setup 1.

\section{Calibration}
\label{sec:calibration}

\subsection{ADC calibration and SiPM cross-talk factor}
\label{sec:x-talk}

Calibration of the ADC response in terms of the number of photons detected by the SiPM was performed by illuminating SiPM with short LED pulses. The driving voltage for the LED had a triangular pulse shape. The amplitude of the driver signal necessary to provide light intensity at which on average 1 photon is detected was 1.1~V. Amplitudes up to 1.22~V were used to create spectra with up to 6 photons on average for the calibration. The width of the driver pulse at the 1~V level was $\sim 2.5\unit{ns}$. The stability of the amplitude was controlled by recording an inverted driver signal in a separate ADC channel.

\begin{figure}
\centering
\includegraphics[width=8.8cm]{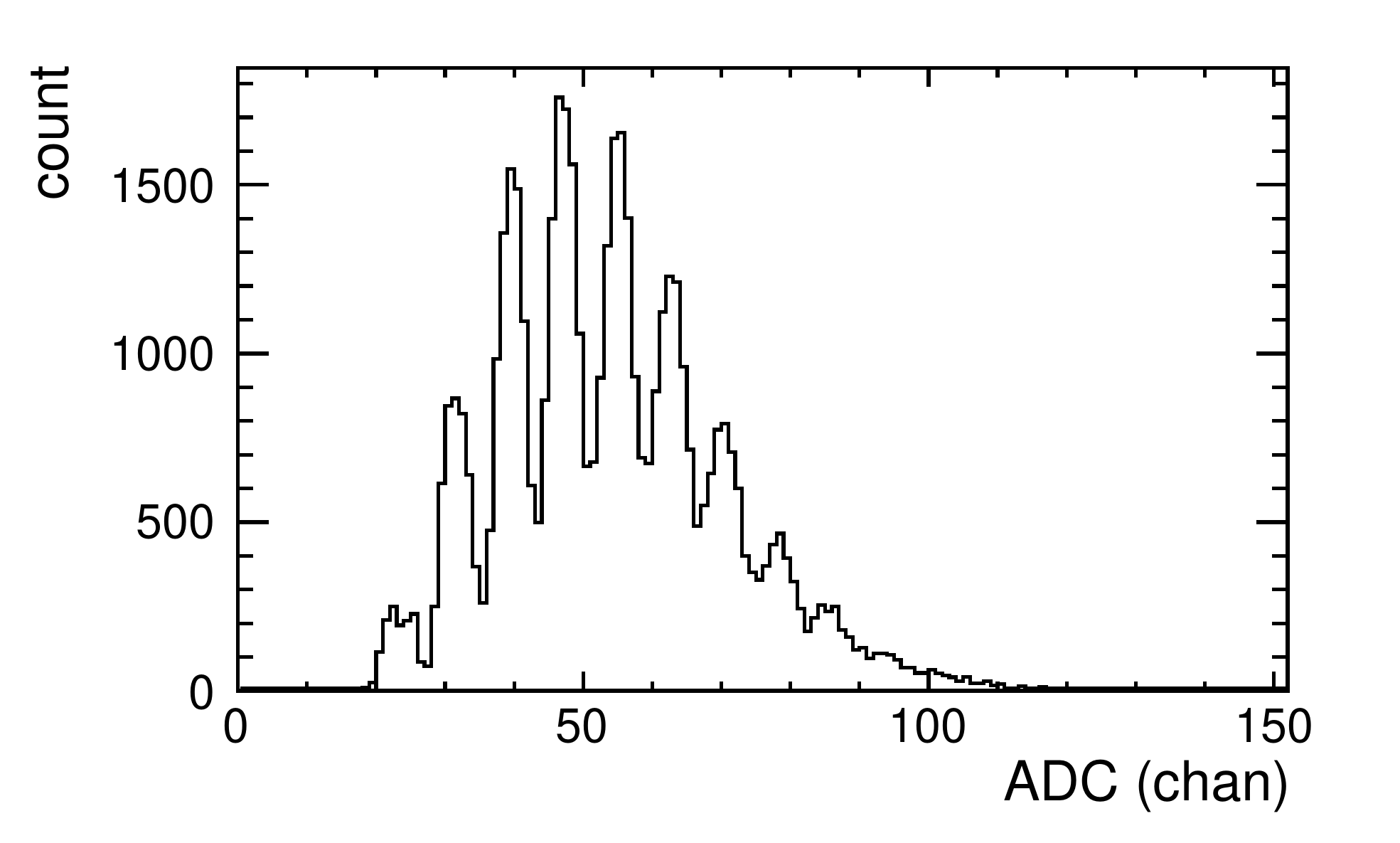}
\caption{\label{fig:photons}ADC spectrum of a SiPM illuminated by LED pulses.}
\end{figure}

Figure \ref{fig:photons} shows an example of the measured ADC spectrum from a SiPM. The peaks in the spectrum correspond to integer numbers of pixels that fire. When the light intensty is low the \emph{pedestal} peak, corresponding to the events in which no pixels have fired, is visible. The centroid of the pedestal peak and the average distance between the centroids of the neighboring peaks are used as calibration constants to express the signal amplitude in terms of the number of pixels that have fired, $n_{pixel}$. 

Each detected photon may cause 1 or more pixels to fire. The ratio of the average number of fired pixels, $\left < n_{pixel} \right >$, to the average number of detected photons, $\nu$, is generally larger than 1 due to the optical and electrical cross-talks, as well as the afterpulsing (See eg. \cite{Dolgo06}). For simplicity we combine all these effects under the common \emph{cross-talk factor} $X = \left < n_{pixel} \right > / \nu$.  

Considering the high number of pixels (3600) in the SiPMs used in the tests, the distribution of the number of detected photons at low light intensity is Poissonian to a very good approximation. Thus the probability for detecting zero photons is $P_0 = e^{-\nu}$. $P_0$ is measured as the ratio of the event number in the pedestal peak, $A_0$, to the integral of the whole spectrum, $A$, allowing to extract the average number of detected photons as $\nu = \ln(A_0 / A)$. On the other hand, the average number of fired pixels is determined from the mean of the measured spectrum. Thus the cross-talk factor is obtained as,

\begin{equation}
\label{eq:x-talk}
X = - \frac{ \left < n_{pixel} \right > }{ \ln(A_0 / A) }
\end{equation}

The cross-talk factor depends on several parameters, including the bias voltage applied to the SiPM and temperature. The bias voltage was kept constant to within 0.1~V for the duration of the tests. The experimental room was climatized, limiting the temperature variations. During the tests, the value of the cross-talk factor was measured several times, and the results were centered around $X = 1.35$ with relative variations up to $\pm 5\%$. 

\subsection{TDC calibration}
\label{sec:tdc-calib}

\begin{figure}
\centering
\includegraphics{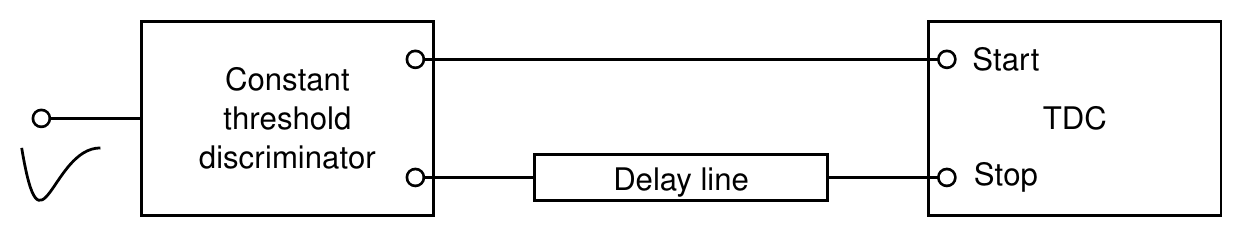}
\caption{\label{fig:tdccalib}Schematic diagram of the circuit used for the TDC calibration.}
\end{figure}

The calibration of the TDC was performed using a setup schematically presented in Fig. \ref{fig:tdccalib}. Pulses from a counter were brought to the input of one of the constant threshold discriminators. One of the logical outputs of the discriminator was used as the start signal for the TDC, while another, after passing through a delay line made of cables of known signal propagation time was used as the stop signal. Delays of 0, 3, 5, 8, 13 and 16~ns were used in addition to the constant difference in cable length between the start and the stop channels. Straight line was fitted to the graph of the delay times vs.\ the centroids of the recorded TDC spectra. The average calibration factor 0.1106~ns/bin was obtained from the slope of the fitted line for the 4 used TDC channels, with relative deviations up to 1\% for the individual channels.

\section{Tested configurations}
\label{sec:configurations}

\begin{figure}
  \centering
  \begin{subfigure}{0.495\textwidth}
    \includegraphics[width=\textwidth]{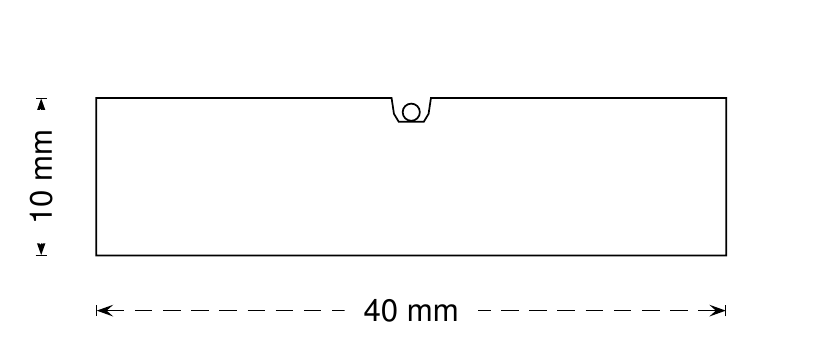}
    \subcaption{\label{fig:confA}}
  \end{subfigure}
  \begin{subfigure}{0.495\textwidth}
    \includegraphics[width=\textwidth]{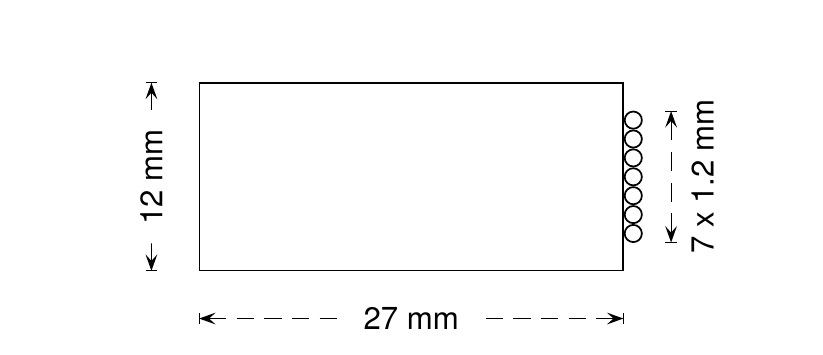}
    \subcaption{\label{fig:confB}}
  \end{subfigure}
  \begin{subfigure}{0.495\textwidth}
    \includegraphics[width=\textwidth]{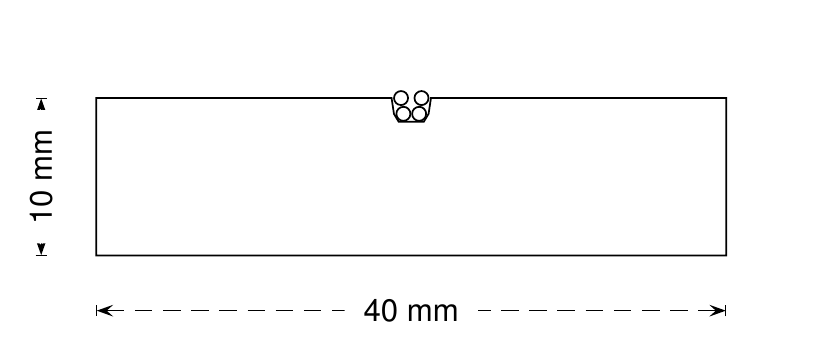}
    \subcaption{\label{fig:confC}}
  \end{subfigure}
  \begin{subfigure}{0.495\textwidth}
    \includegraphics[width=\textwidth]{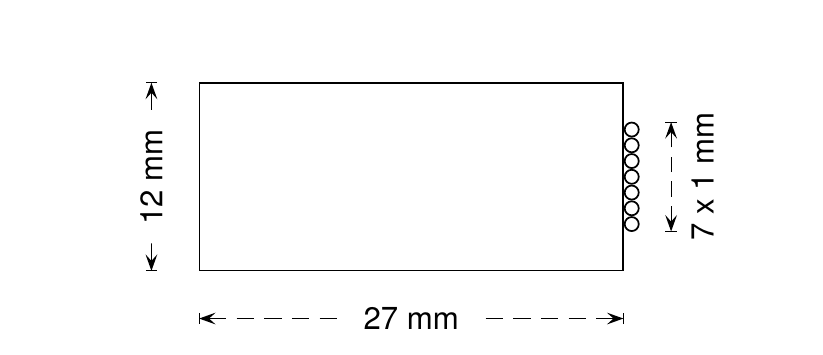}
    \subcaption{\label{fig:confD}}
  \end{subfigure}
  \caption{\label{fig:configurations} Tested configurations of 
                  the scintillator strip with WLS fibers: \\
                  \subref{fig:confA} \cfa -- 
                         MINOS strip with one Kuraray WLS fiber,\\
                  \subref{fig:confB} \cfb -- 
                         Bicron strip with 7 Kuraray WLS fibers,\\
                  \subref{fig:confC} \cfc -- 
                         MINOS strip with 4 Bicron WLS fibers,\\
                  \subref{fig:confD} \cfd -- 
                         Bicron strip with 7 Bicron WLS fibers.
          }
\end{figure}

Four configurations of the scintillator strip with WLS fibers were tested for the light yield. The description of the four configurations, denoted $A$, $B$, $C$ and $D$, is given in the following subsections and schematically presented in Fig. \ref{fig:configurations}. The configurations were designed based on the experience from the upgrade of the \dzero muon system  \cite{Evdoki98}. Two configurations with the best light yield were selected for detailed position- and time-resolution tests.

\subsection{\Cfa}

\Cfa consisted of a 1~m long clear polystyrene scintillator strip with a $41\times10\unit{mm}^2$ profile, co-extruded with a $\text{TiO}_2$ loaded polystyrene layer such as used by the MINOS collaboration \cite{MINOS_sci_08}, with one polystyrene double-clad fiber of 1.2~mm diameter with 175 ppm of Y-11 fluor produced by Kuraray Inc.\ Japan, inserted into the groove (Fig. \ref{fig:confA}) \cite{Kuraray}. The strip is then wrapped in black paper. The average light yield in \cfa was measured to be 10 photoelectrons on each side of the strip. 

\subsection{\Cfb}

\Cfb consisted of a 1~m long clear \bicron fast scintillator strip with a $27\times12\unit{mm}^2$ profile \cite{BicronStrip}, with 7 \kuraray Y-11 WLS fibers arranged along the narrow side (Fig. \ref{fig:confB}). The fibers were attached to the strip in several points along the strip using reflective tape. The configuration was then wrapped with one layer of white Tyvek, and two layers of black paper.

The average light yield in the \cfb was measured to be 19 photoelectrons on each side. 

\subsection{\Cfc}

\Cfc consisted of the MINOS strip \cite{MINOS_sci_08}, with 4 \bicronwls WLS fibers of 1.0~mm diameter inserted into the groove (Fig. \ref{fig:confC}) \cite{BicronWLS}. The average light yield in \cfc was measured to be 20 photoelectrons on each side.

\subsection{\Cfd}

\Cfd consisted of the \bicron strip \cite{BicronStrip}, with 7 \bicronwls WLS fibers of 1.0~mm diameter each arranged along the narrow side of the strip (Fig. \ref{fig:confD}). The average light yield in \cfd was measured to be $\sim 30$ photoelectrons on each side. 

The configurations $C$ and $D$ were selected for detailed position- and time-resolution studies.

\section{Analysis}
\label{sec:analysis}

\subsection{Analysis of the measurements performed with setup 1}

\subsubsection{Event selection}

For setup 1 the off-line event selection was performed as follows:

\begin{enumerate}

  \item Events in which the \sfour signal is below the discriminator threshold, signalled by the end-of-scale value in the respective TDC channel, were rejected. 

  \item An amplitude threshold was imposed on the \sfour counter signals to reject events with energies below the characteristic Landau distribution for the muon energy deposition. 

  \item Events in which the time of \sipmo or \sipmt shows the TDC end-of-scale value were rejected. 

  \item On rare occasions, the noise in the counter signals causes premature threshold crossing in the discriminator. Such events are characterized by a zero TDC readout in the respective channel, and were rejected. The fraction of such events was below 1\% in all analyzed data samples.

\end{enumerate}

\subsubsection{Position resolution and the speed of signal propagation along the strip}
\label{sec:position1}

In setup 1, the location of the muon impact was defined by the position $x$ of the \sfour counter along the axis of  the tested strip. The center of the tested strip was assigned the relative position $x=0$, and the $x$ axis was oriented away from \sipmt towards \sipmo. Five points along the strip were measured for the configurations $C$ and $D$. The data for each point were collected over 7 to 12 hours in order to collect sufficient statistics (at least 500 events remaining per point after the selection cuts).

The observable with the best sensitivity to muon position along the strip is the time difference between the two SiPMs. Position is thus measured as,

\begin{equation}
\label{eq:sipmx}
   x_{SiPM} = b_0 + v \frac{t_{SiPM_2} - t_{SiPM_1}}{2} = b_0 + v \frac{\Delta t}{2}
\end{equation}

where $b_0$ is the offset and $v$ is the speed of the signal propagation along the tested strip. 

\begin{figure}
  \centering
  \begin{subfigure}[b]{0.495\textwidth}
    \includegraphics[width=\textwidth]{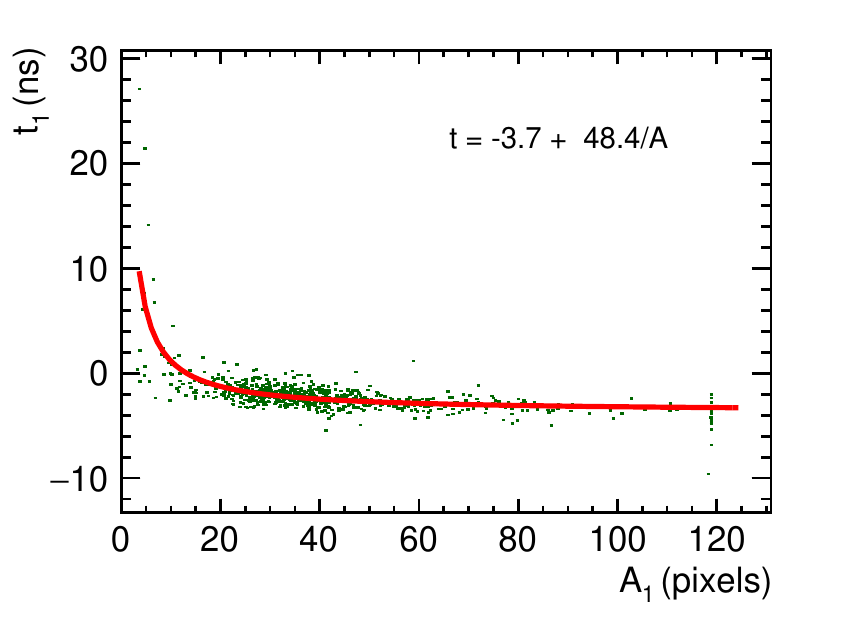}
    \subcaption{\label{fig:aCorr1}}
  \end{subfigure}
  \begin{subfigure}[b]{0.495\textwidth}
    \includegraphics[width=\textwidth]{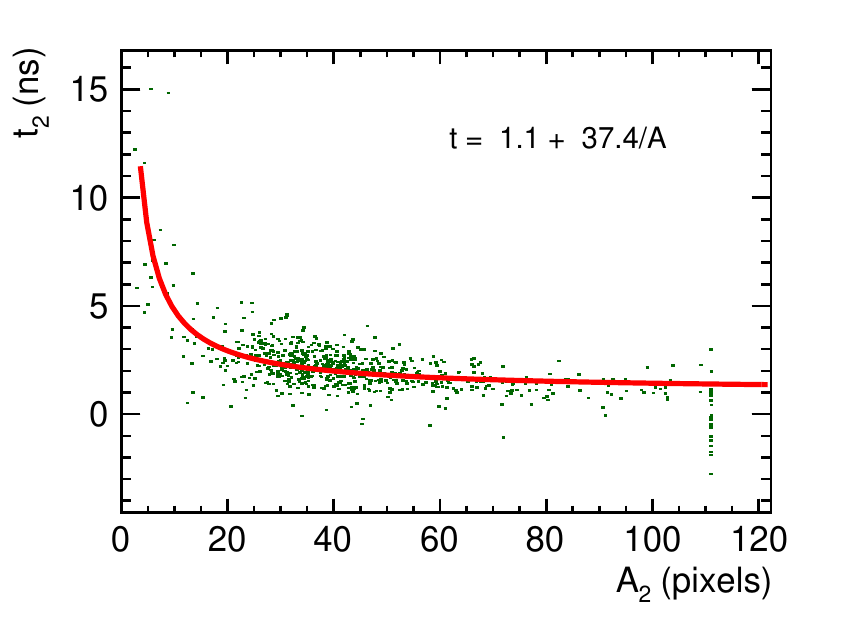}
    \subcaption{\label{fig:aCorr2}}
  \end{subfigure}
  \caption{\label{fig:aCorr} Scatter plot of time vs.\ amplitude of \sipmo 
                  \subref{fig:aCorr1} and \sipmt \subref{fig:aCorr2}, from the
                  dataset taken at the \sfour position $x=0$ with \cfd, showing 
                  the amplitude-dependent delay of the timing signal from the 
                  discriminator. The function used for the correction 
                  of this effect is also shown (red line).}
\end{figure}

Timing of the signal rising edge using constant threshold discriminator has well-known amplitude-dependent variations. In the tested strips, the amplitude effect, beside worsening the time resolution also introduces a position-dependent bias because of the attenuation of light along the strip. The amplitude effect was corrected by fitting the function of the form $t = a_0 + a_1/A$, where $A$ is the amplitude of the signal, to the scatter plot of time versus amplitude for each SiPM separately (Fig. \ref{fig:aCorr}). The uncertainty of the amplitude was neglected in the fit. The parameter $a_1$ obtained in the fit to the data collected at $x=0$ (\sfour at the center of the tested strip) was used to correct the amplitude effect at all measured positions for the same tested configuration.

\begin{figure}
\centering
\includegraphics{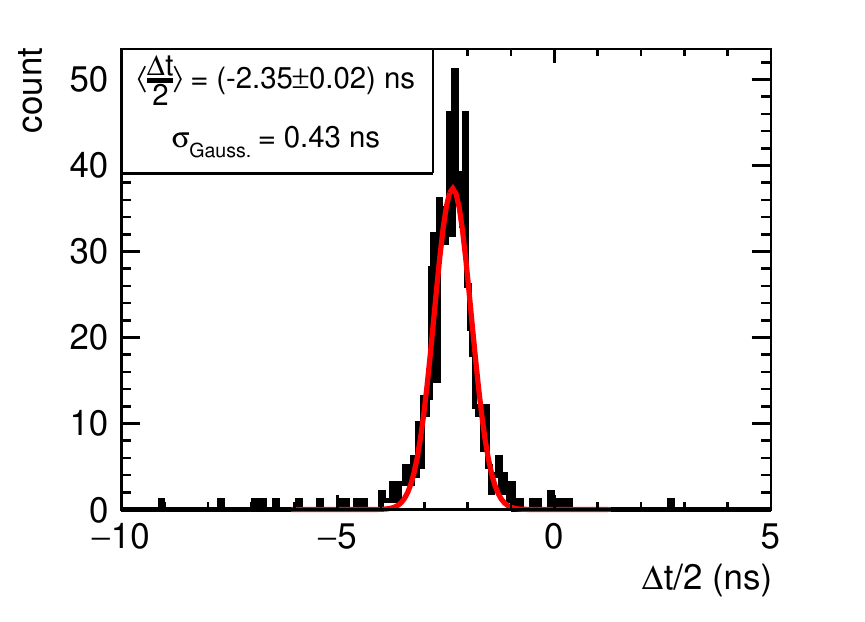}
\caption{\label{fig:x0}Distribution of the variable $\Delta t/2$ after the amplitude correction for the \cfd and $x=0$. }
\end{figure}

The distribution of the variable $\Delta t/2$ after the amplitude correction is shown in Fig. \ref{fig:x0} for the \cfd and $x=0$. The standard deviation of the distribution, determined from a Gaussian fit, is the crucial parameter for the counter position resolution. 

\begin{figure}
\centering
\includegraphics{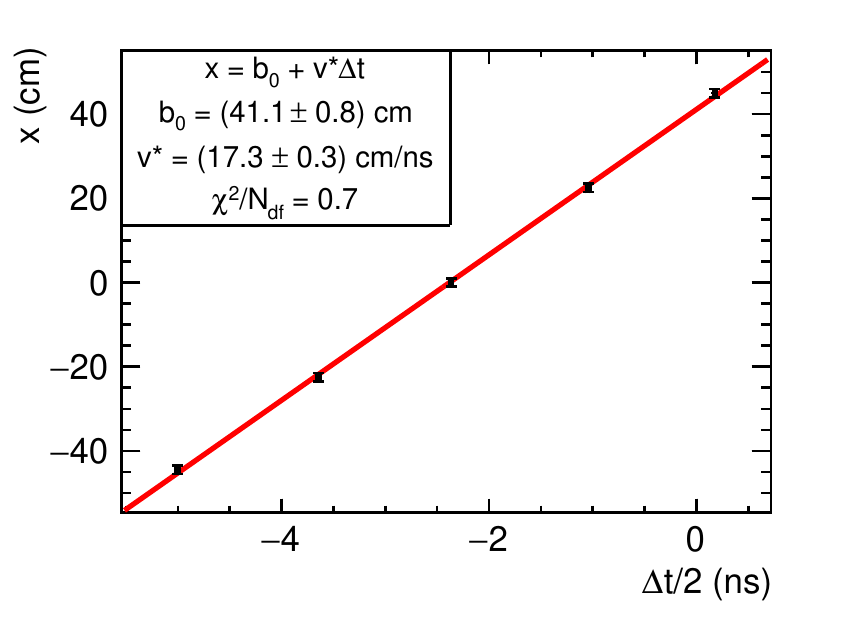}
\caption{\label{fig:speed1} Scatter plot of $x$ vs. $\frac{\Delta t}{2}$ for the \cfd. }
\end{figure}

The speed of the signal propagation along the strip is determined from the linear fit to the scatter plot of $x$ vs. $\Delta t/2$, as shown in Fig. \ref{fig:speed1}. The uncertainty on $\Delta t/2$ was estimated from the scatter of the data, while the uncertainty on $x$ was set to 1~cm, corresponding to the estimated precision of the position of the \sfour counter. 

The fitted value of the signal propagation speed along the strip in this example was 17.3~cm/ns. The average value of the standard deviation of $\Delta t/2$ from all 5 measured points was $\sigma_{\Delta t/2} = 0.45 \unit{ns}$, which directly translates into the estimated position resolution $\sigma_x = 7.8\unit{cm}$ for the \cfd. Beside the position resolution of the tested strip, this estimate contains contributions from the uncertainty of the muon impact position along the tested strip due to the width of \sfour and the uncertainty due to the angular distribution of the muon tracks across the distance between \sfour and the tested strip. This estimate of the coordinate resolution along the tested counter can thus be regarded as conservative.

\subsubsection{Time resolution}
\label{sec:time1}

\begin{figure}
\centering
  \begin{subfigure}[b]{0.495\textwidth}
  \includegraphics[width=\textwidth]{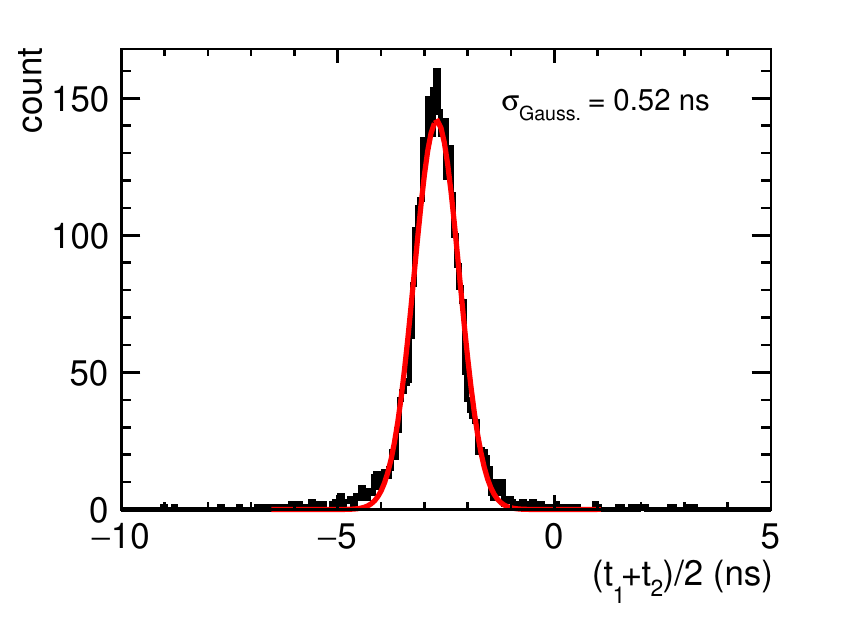}
    \subcaption{\label{fig:time1-plus}}
  \end{subfigure}
  \begin{subfigure}[b]{0.495\textwidth}
  \includegraphics[width=\textwidth]{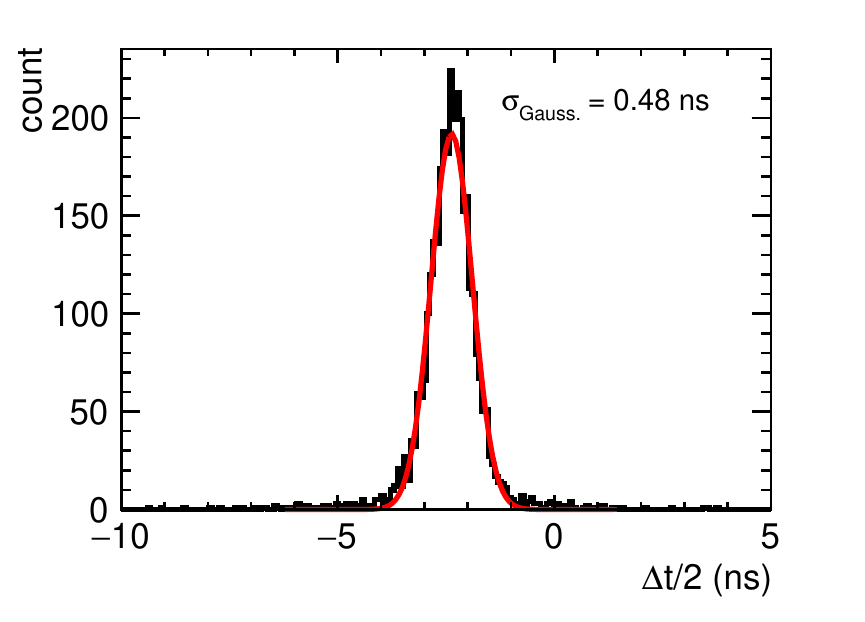}
    \subcaption{\label{fig:time1-minus}}
  \end{subfigure}
  \caption{\label{fig:time1} Distribution of the mean times of 
              the strip \cfd w.r.t.\ \sfour in setup 1 \subref{fig:time1-plus}
              and distribution of the time difference between the two SiPM, 
              plotted with 
              correction for the muon position \subref{fig:time1-minus}. 
              Gaussian fit is shown as the red line in both plots.}
\end{figure}

The most direct way to measure strip time resolution is to analyze the 
distribution of differences between the mean time of the tested strip and
the time of \sfour. As the mean time of the tested strip does not depend on 
the position of \sfour along the tested strip, the data from all 5 studied 
\sfour positions can be added up.
The distribution of mean times is shown in Fig. 
\ref{fig:time1-plus} for the \cfd. The standard deviation of the fitted 
Gaussian curve is 0.52~ns. In this plot, the amplitude effect correction was 
applied to \sfour in a similar way as to the SiPM. 

The dominant contribution to the width of the distribution is the time 
resolution of the tested strip, but other contributions are also present, 
such as the time resolution of \sfour. Also, the width of the tested strip 
introduces a spread in the muon positions w.r.t.\ the PMT of \sfour.
Another estimate of the time resolution of the tested strip can be 
inferred from the width of the distribution of $\Delta t/2 
= (t_{\sipmo} - t_{\sipmt})/2$. A plot of $\Delta t/2 - x/v$ from all five datasets
corresponding to different positions $x$ of \sfour, is shown in Fig.~\ref{fig:time1-minus} for the \cfd. The subtraction of the signal propagation delay $x/v$ 
takes care of the dependence of $\Delta t/2$ on $x$. The standard deviation 
of the fitted Gaussian curve in this plot is 0.48~ns. The time resolution of 
\sfour does not 
contribute to this width. Nevertheless, contributions 
from uncertainties other than the time resolution of the tested strip are 
still present, such as the uncertainties of $x$ and $v$, and the muon-position
spread due to the width of \sfour.

\subsection{Analysis of measurements performed in setup 2}

In setup 2 the position of the muon hit is determined from the time difference between \sthrt and \sthro, $\Delta t_{\sthr} = t_{\sthrt} - t_{\sthro}$ in the same way as for the tested strips (Eq. \ref{eq:sipmx}). The mean effective speed of the signal propagation along the scintillator strip, $v_{sc}$, is generally lower than the speed of light in the medium itself due to the light travelling under non-zero angle w.r.t.\ the strip and reflecting back into the medium multiple times before reaching the photo detector. 

\begin{figure}
\centering
\includegraphics{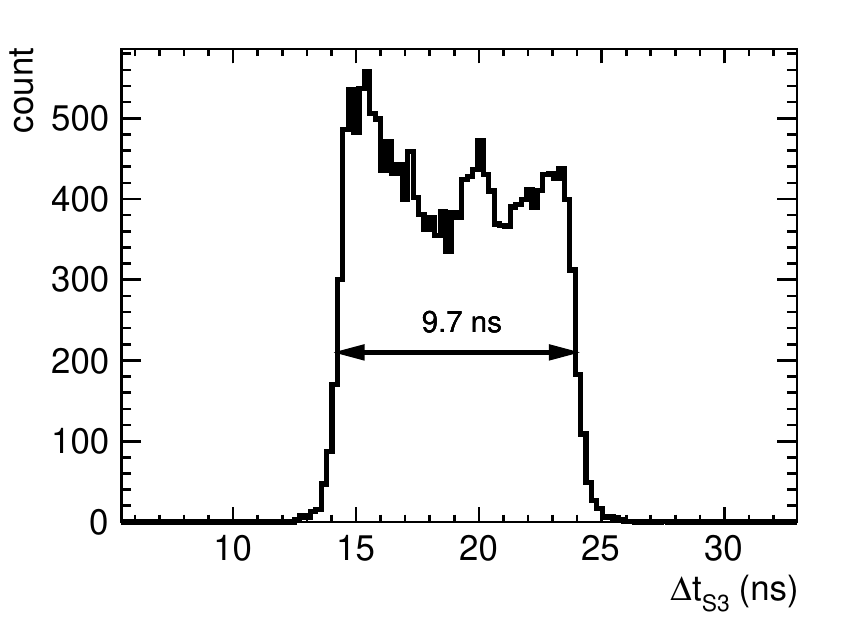}
\caption{\label{fig:tdc4}Time distribution of \sthrt. The structure of the distribution reflects the variation of detection efficiency for cosmic muons along the setup.}
\end{figure}

An example of the time distribution of the \sthrt channel is shown in Fig.~\ref{fig:tdc4}. The distribution has a structured shape with a width of 9.7~ns, and a Gaussian smearing due to the time resolution of \sthro and \sthrt. The observed shape is mainly due to non-uniform detection efficiency of the system of counters. Calibration of the offset and the signal propagation speed is achieved by assigning the time at half maximum at the high-$\Delta t$ edge of the distribution to the \sthro end, and half-maximum at the low-$\Delta t$ edge to the \sthrt end. The length of the \sthr scintillator in this test was $l = 60 \unit{cm}$, leading to $v_{sc} = 12.4 \unit{cm/ns}$. 

\subsubsection{Event selection}

In setup 2 the off-line event selection was performed as follows:

\begin{enumerate}

  \item An amplitude threshold was imposed on the sum of amplitudes of \sthro and \sthrt to reject events at energies below the characteristic Landau distribution for the muon energy deposit. 

  \item Events in which the time of \sipmo, \sipmt or \sthrt shows the TDC end-of-scale value were rejected. 

  \item Events with zero TDC reading in any of the channels were rejected. The fraction of such events was below 1\% in all analyzed data sets.

\end{enumerate}

\subsubsection{Attenuation of light along strips with WLS fibers}
\label{sec:attenuation}

\begin{figure}
\centering
   \includegraphics[width=8.8cm]{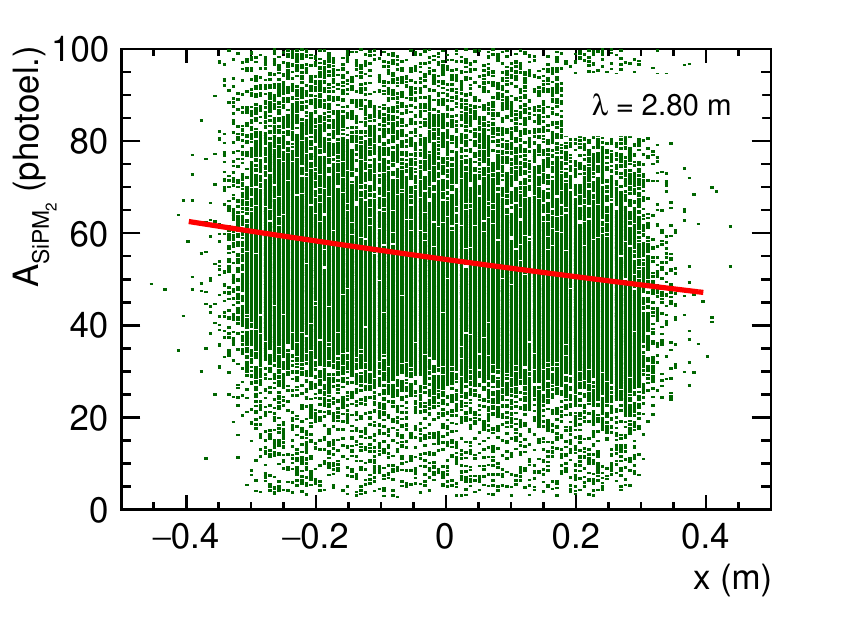}
   \caption{\label{fig:attenuation} 2D plot of the \sipmt amplitude vs.\ position event distribution along the \sthr strip. Fit of the exponential function and the attenuation length are also shown.}
\end{figure}

Figure \ref{fig:attenuation} shows a 2D plot of the \sipmo amplitude vs.\ position along the \sthr strip in the \cfd. Exponential fit to the data indicates an attenuation length for the light signal in this strip of $\lambda = 2.8 \unit{m}$. The light intensity is well preserved by the WLS fibers along the strip, so muon position cannot be precisely determined from the signal attenuation.

\subsubsection{Position resolution from $\Delta t$ and the speed of signal propagation along the strip}
\label{sec:position2}

\begin{figure}
  \centering
  \begin{subfigure}[b]{0.495\textwidth}
    \includegraphics[width=\textwidth]{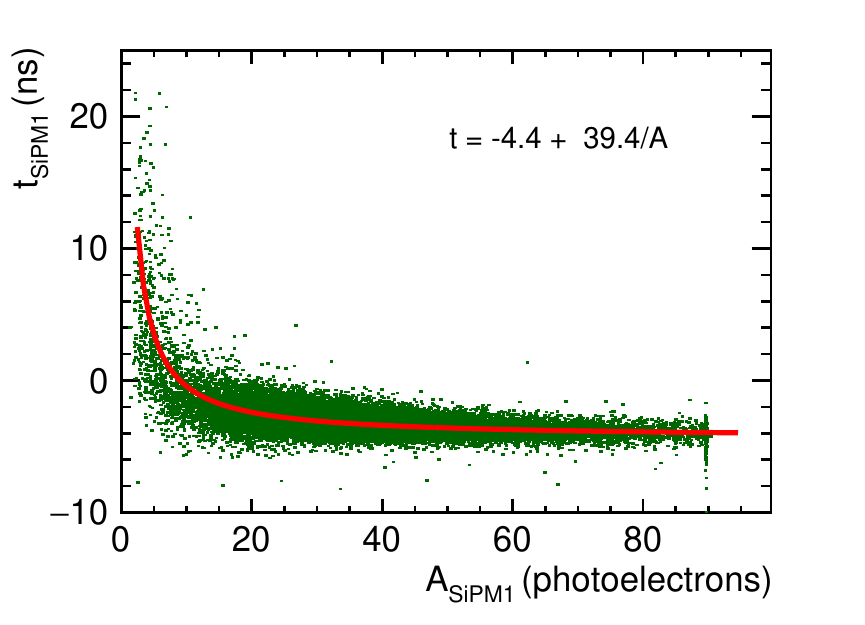}
    \subcaption{\label{fig:aCorr12}}
  \end{subfigure}
  \begin{subfigure}[b]{0.495\textwidth}
    \includegraphics[width=\textwidth]{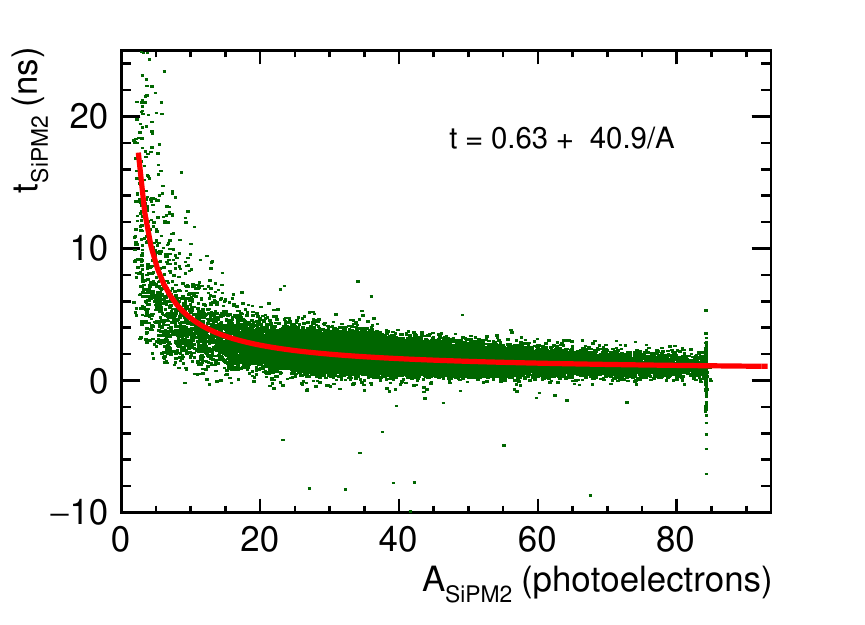}
    \subcaption{\label{fig:aCorr22}}
  \end{subfigure}
  \caption{\label{fig:thrdelay} Scatter plot of time vs.\ amplitude of \sipmo 
                  \subref{fig:aCorr12} and \sipmt \subref{fig:aCorr22}, from the
                  dataset taken with the \cfd, showing 
                  the amplitude-dependent delay of the timing signal from the 
                  discriminator. The function used for the correction 
                  of this efect (see legend) is also shown.}
\end{figure}

The amplitude effect for setup 2 was corrected in a similar way as for setup 1. Figure \ref{fig:thrdelay} shows the scatter plot of time versus amplitude of both SiPMs from the dataset taken with the \cfd. The fit of the function $t = a_0 + a_1/A$ is also shown. The uncertainty on the amplitude was neglected in the fit. 

\begin{figure}
\centering
  \begin{subfigure}[b]{0.495\textwidth}
     \includegraphics[width=\textwidth]{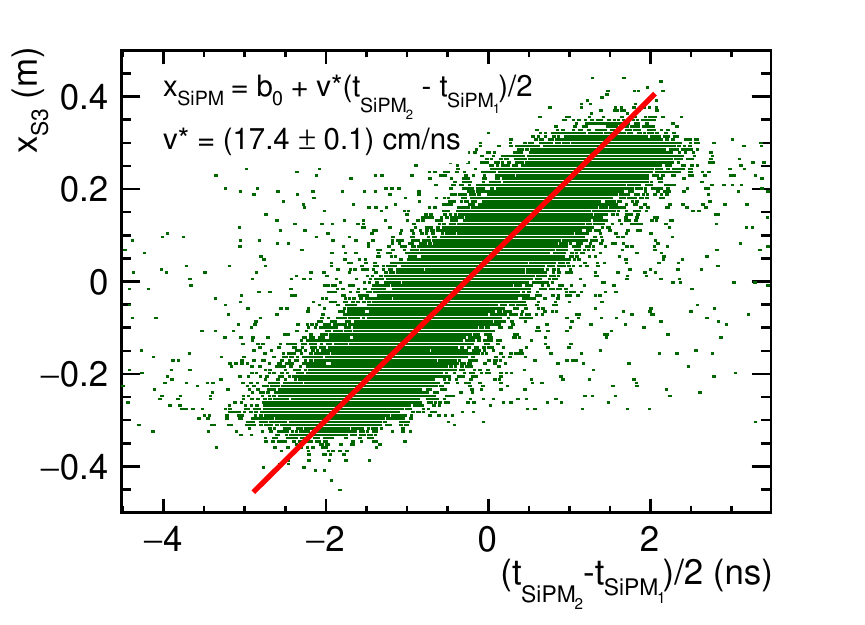}
     \subcaption{\label{fig:xcalib} }
  \end{subfigure}
  \begin{subfigure}[b]{0.495\textwidth}
     \includegraphics[width=\textwidth]{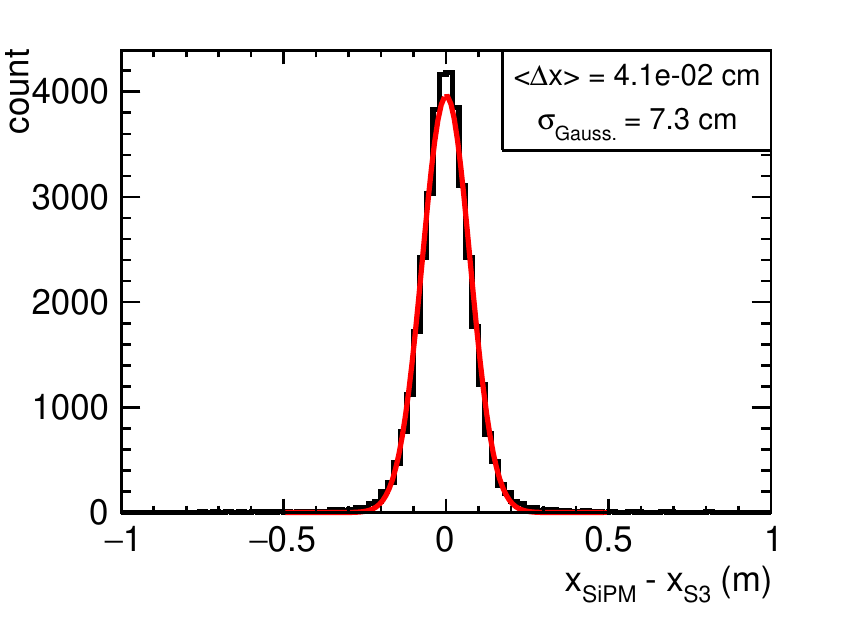}
     \subcaption{\label{fig:xdev}}
  \end{subfigure}
  \caption{\subref{fig:xcalib} Scatter plot of the position along the strip 
              measured with \sthr vs.\ $(t_{\text{SiPM}_2} - t_{\text{SiPM}_1})/2$
              in the \bicron configuration. Linear fit to the data is also shown. 
           \subref{fig:xdev} Distribution of deviations of $x_{\text{SiPM}}$ w.r.t\ $x_{\text{S3}}$.}
\end{figure}

Figure \ref{fig:xcalib} shows the scatter plot of the position $x_{\text{S3}}$ measured with \sthr vs.\ $(t_{\text{SiPM}_2} - t_{\text{SiPM}_1})/2$, after amplitude correction. Position $x_{\text{S3}}$ is determined from $t_{\text{S32}}$ assuming that the width at half maximum of the $t_{\text{S32}}$ distribution (see Fig. \ref{fig:tdc4}) corresponds to the full length of the \sthr strip. 

The speed of the signal propagation along the strip determined from the linear fit to the scatter plot in this example is in excellent agreement with the result obtained with setup 1. The uncertainty of the position measurement is estimated from the distribution of the deviation of the position measured with SiPM w.r.t\ the position measured with \sthr on the same dataset (Fig. \ref{fig:xdev}). The standard deviation estimated from the fitted Gaussian curve is $\sigma_{\text{Gauss.}} = 7.3 \unit{cm}$. This includes the position uncertainty of the tested strip, in addition to the position uncertainty of the reference strip, as well as the effect of the angular distribution of muons across the distance between the strip centers. It thus represents a conservative estimate of the position resolution of the tested strip.

\subsubsection{Time resolution}
\label{sec:time2}

\begin{figure}
\centering
  \includegraphics{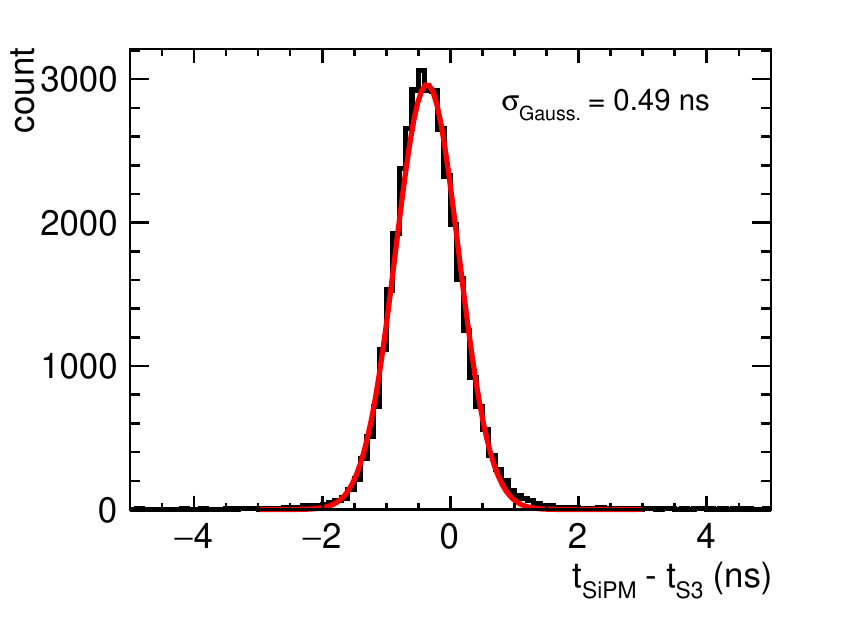}
  \caption{\label{fig:meantime} Distribution of the average-time differences 
              between the tested strip and \sthr. Gaussian fit is also shown.}
\end{figure}

The time resolution of the tested strip is directly measured by subtracting the average time of \sthro and \sthrt from the average time of \sipmo and \sipmt. The plot of the difference is shown in Fig.\ \ref{fig:meantime}. The standard deviation of the distribution extracted from the Gaussian fit is 0.49~ns.

\subsubsection{SiPM time resolution}
\label{sec:time2a}

For completeness, an estimate of the time resolution of the individual SiPM readouts is made as follows.

The \sthr phototubes were used as the time reference for the respective SiPM signals. The time of \sthro was subtracted from the time of \sipmo, and \sthrt was subtracted from \sipmt as a first step of the analysis. Subsequently, the effect of the difference in the signal propagation speeds in \sthr and the tested strip was corrected based on the position $x$ measured by \sthr. The amplitude effect was corrected as described in  Sec. \ref{sec:position2}.

\begin{figure}
\centering
  \includegraphics{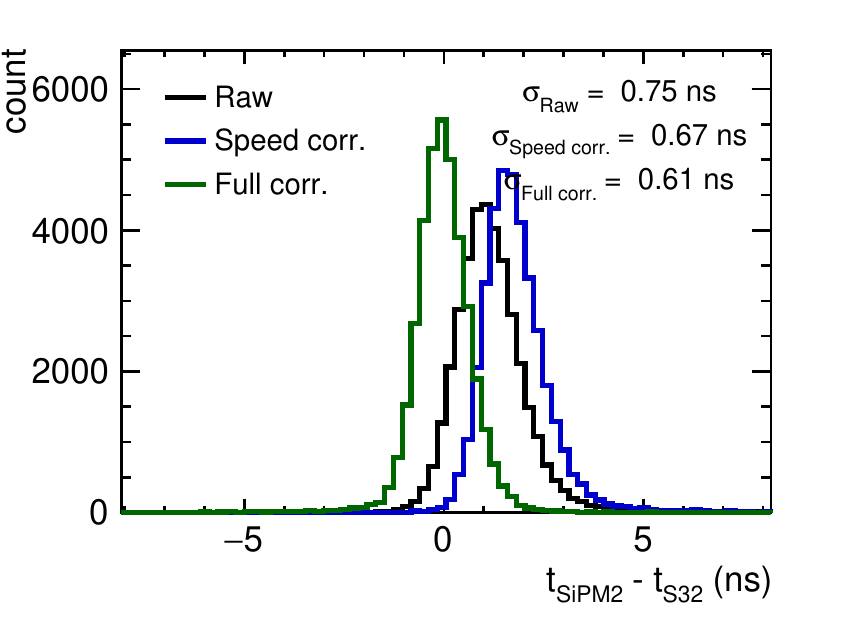}
  \caption{\label{fig:resolution} Time histogram for one of the SiPM channels
                  w.r.t.\ the reference counter, without corrections (black), 
                  with correction for the difference in the signal propagation 
                  speed in the strips (blue) and with full correction including 
                  amplitude corrections for the tested strip and the 
                  reference strip (green).}
\end{figure}


The time resolution of the \sipmt in the \cfd is shown in Fig. \ref{fig:resolution} for the raw case (without any corrections, black line), after correction for the difference in the signal propagation speed in the strips (blue line) and after the amplitude correction (green line). The numerical values shown in the figure represent fitted Gaussian standard deviations of the respective measured distributions (curves not shown).

\section{Results}
\label{sec:results}

Table \ref{tab:sipm} summarizes the measured performance of the individual SiPM readout channels for the configurations $C$ and $D$. 

Table \ref{tab:results} summarizes the results regarding the strip configurations as a whole. The difference in the resolutions for the two configurations is larger than the ratio of the statistical factors corresponding to the photon yield per muon in each configuration. This indicates that other factors influence the time and the position resolutions besides the statistical effect of the photon yield. These additional effects include the properties of the scintillator and the WLS fiber materials such as the light emission time, the geometry of the configuration, the location of the WLS fibers etc. 

The best directly measured position resolution is 7.3~cm, and has been achieved with the \cfd. 

\begin{table}
\caption{\label{tab:sipm} Measured performance of the individual SiPM readout 
   channels for the configurations $C$ and $D$.}
\centering
	\begin{tabular}{ c | c | c | c }
    \hline
	Configuration  & SiPM  & Light yield / muon & $\sigma_{t,SiPM}$ \\
		           &   \#  &  (photoelectrons)  &  (ns)  \\
	\hline
	\multirow{2}{*}{$C$} & 1 & 21 & 1.12 \\
                         & 2 & 20 & 1.17  \\
    \hline
	\multirow{2}{*}{$D$} & 1 & 31 & 0.67 \\
	                     & 2 & 36 & 0.61 \\
    \hline
	\end{tabular}
\end{table}

\begin{table}
\caption{\label{tab:results} Measured properties of the strip configurations 
        $C$ and $D$.}
\centering
	\begin{tabular}{ c | c | c | c | c  }
    \hline
	Configuration  & $\sigma_x$ (setup 1) & $\sigma_x$ (setup 2) & $\sigma_{t,strip}$ & 
                     Speed of light \\
		           &   (cm)               &     (cm)             & (ns)       & 
                        (cm/ns)  \\
	\hline
	$C$            &   14.8  &  14.8  &  0.91  &  18.1 \\
	$D$            &   7.8   &  7.3   &  0.48  &  17.2 \\
    \hline
	\end{tabular}
\end{table}

\section{Conclusions}
\label{sec:conclusions}

Prototype scintilator+WLS strip configurations with SiPM readout for the muon system for future colliders were tested for light yield, position resolution and time resolution. Depending on the configuration, a light yield in single SiPM of up to 36 photoelectrons per muon has been achieved. Two different setups were used for position- and time-resolution measurements. In one setup, the muon impact position was determined by coincidence requirement with a small-area counter. In the other, a reference strip with vacuum PMT parallel to the tested strip was used to measure position independently.

Strip time resolution of \besttime and position resolution of \bestpos were achieved. Tests with more precise timing and/or position reference, such as in the test-beam, will yield results with higher precision.

\printbibliography[title=References]

\end{document}